\documentstyle[prc,preprint,epsf,rotate,aps]{revtex}
\begin{document} 
\tighten
\draft 
%\preprint{ } 
\title{Pairing and symmetry energy in $N \simeq Z$ nuclei} 
\author{ P. Vogel}  
\address{ Physics Department 161-33, California
Institute of Technology, Pasadena, California 91125  
} 
\date{\today} 
\maketitle

\begin{abstract} 
It is shown that in $N \simeq Z$ nuclei the effects of isovector pairing and
the symmetry energy are intimately related. In particular, in the odd-odd 
$N = Z$ nuclei these two effect are essentially equal in magnitude, causing
near degeneracy of the lowest $T = 1$ and $T = 0$ states and appearance
of the $T = 1$ ground state isospin in many such nuclei. 
Following the earlier work of J\"{a}necke, it is shown that the global
symmetry energy fit can be reproduced using just the excitation energies
of the lowest $T = 3/2$ states in $N = Z+1$ nuclei. Similarly,
the global pairing gap fit can be related to the excitation energy
of the $T = |N - Z|/2 +1$ states in $N \simeq Z$ even $A$ nuclei.

\end{abstract} 

The treatment of pairing correlations in proton rich $N \simeq Z$ nuclei,
where the neutron-proton ($np$) pairing is expected to play an important role,
became a popular subject recently 
\cite{ELV,Langanke96,Wyss,Civitarese0,Civitarese,Dobes,Poves}.  
Arguments based on isospin symmetry suggest that the isovector
$np$, $J^{\pi} = 0^+$ pairing should be considered on the same footing 
as the conventional like-particle
$nn$ and $pp$, $J^{\pi} = 0^+$ pairing correlations. 
In addition, neutrons and protons can
also interact via an isoscalar force; this new mode of pairing
with $J \ne 0$ has been also recently extensively discussed
\cite{SO8,Piet}.

\begin{figure}
\begin{center}
\leavevmode 
\rotate[r]{\epsfxsize=0.7\textwidth 
\epsffile{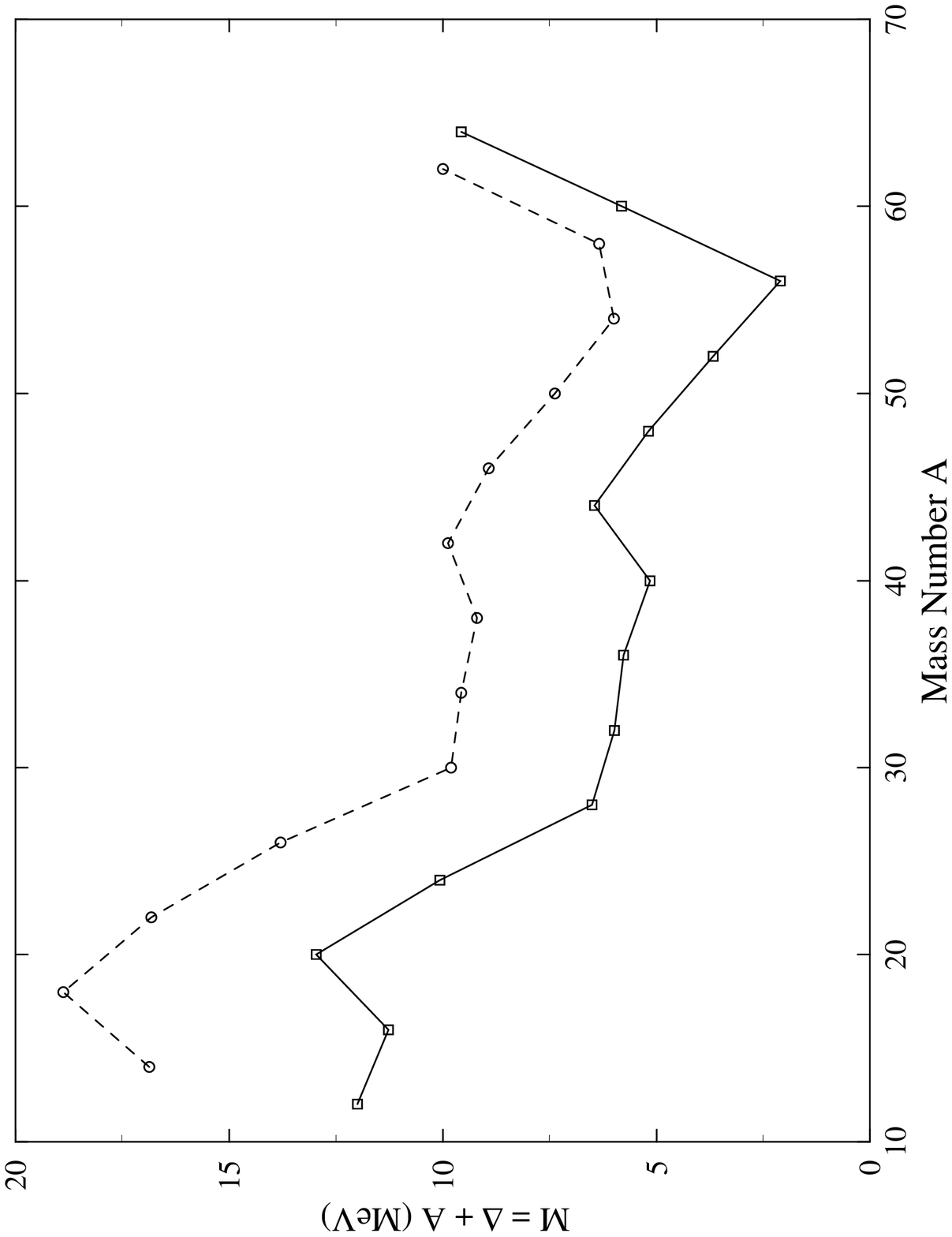}}
%\epsffile{eeoo2.ps}}
\vspace{1.5cm}
\caption{Mass excess of the $N = Z$ nuclei. 
    The dashed line with circles connects the masses of the odd-odd nuclei 
    and the full line with squares the even-even nuclei.  For easier viewing,
    the $\Delta + A$ instead of $\Delta$ is plotted versus $A$. 
}
\label{f:bind}
\end{center}
\end{figure}

The obvious experimental
signature of pairing correlations is the extra binding
energy of the even-even nuclei relative to their odd-A neighbors.
(This feature is present even in the more general situation
as shown in Ref. \cite{Jacek}.) 
Moreover, in stable nuclei with $N > Z$,  odd-odd nuclei are 
even less bound 
than their odd-A counterparts. Now, if pairing correlations indeed have
an isovector character, then the odd-odd $N = Z$ nuclei 
should be as fully paired 
as the even-even $N = Z$ nuclei are and
one could therefore expect that their masses form a single smooth line
as a function of the mass number $A$.
However, as shown in Fig. \ref{f:bind},
the experiment does not support this assumption. 
In fact, the opposite is true: as one can see 
the odd-odd nuclei are systematically less
bound than the even-even ones. Does it mean that the idea of isovector
pairing is incorrect and the $np$ pairing is absent?
The answer is no. 
In this paper I explain this seemingly paradoxical
situation, and show its relation to the interplay 
between the pairing and symmetry
energies. This is a phenomenological analysis based on the experimental
nuclear masses and excitation energies. No attempt is made to relate
the facts to the underlying many-body theory. Instead, the smooth
trends are revealed and categorized.
I concentrate on the effects associated with the isovector pairing; 
it is assumed that the isoscalar pairing is included as a component
of the nuclear symmetry energy.

The considerations below follow the much earlier work of
J\"{a}necke \cite{Janecke} which, in turn, has been apparently in part
inspired by an even older work of Baz and Smorodinsky \cite{Baz}.
In Ref. \cite{Janecke} the empirical values of the symmetry and
pairing energies were deduced from the excitation energies
of the analog states. These then could be used in order to predict
the masses of unknown nuclei. Our motivation here is different.
Nevertheless, there is a close connection with the work of J\"{a}necke
\cite{Janecke}, as shown throughout the text below.

The clue to the paradox in Fig.\ref{f:bind} is the isospin.
All even-even $N = Z$ nuclei have $T = 0$ ground states, 
while the odd-odd $N = Z$ nuclei have either $T=1$ ground state or the
$T = 1$ state is one of the low-lying states in the spectrum. It is well known 
\cite{BM} that
nuclear binding depends on the smooth volume, surface, and Coulomb
energies and, in addition, on the symmetry energy which  reduces 
the binding energy of the $T = 1$ states  
with respect to the $T = 0$ states, explaining the trend in Fig.\ref{f:bind}. 
Thus, in the $N \simeq Z$ nuclei one must not
treat the pairing alone; the effect of the symmetry energy should be 
included as well.

\begin{figure}
\begin{center}
\leavevmode 
\rotate[r]{\epsfxsize=0.7\textwidth 
\epsffile{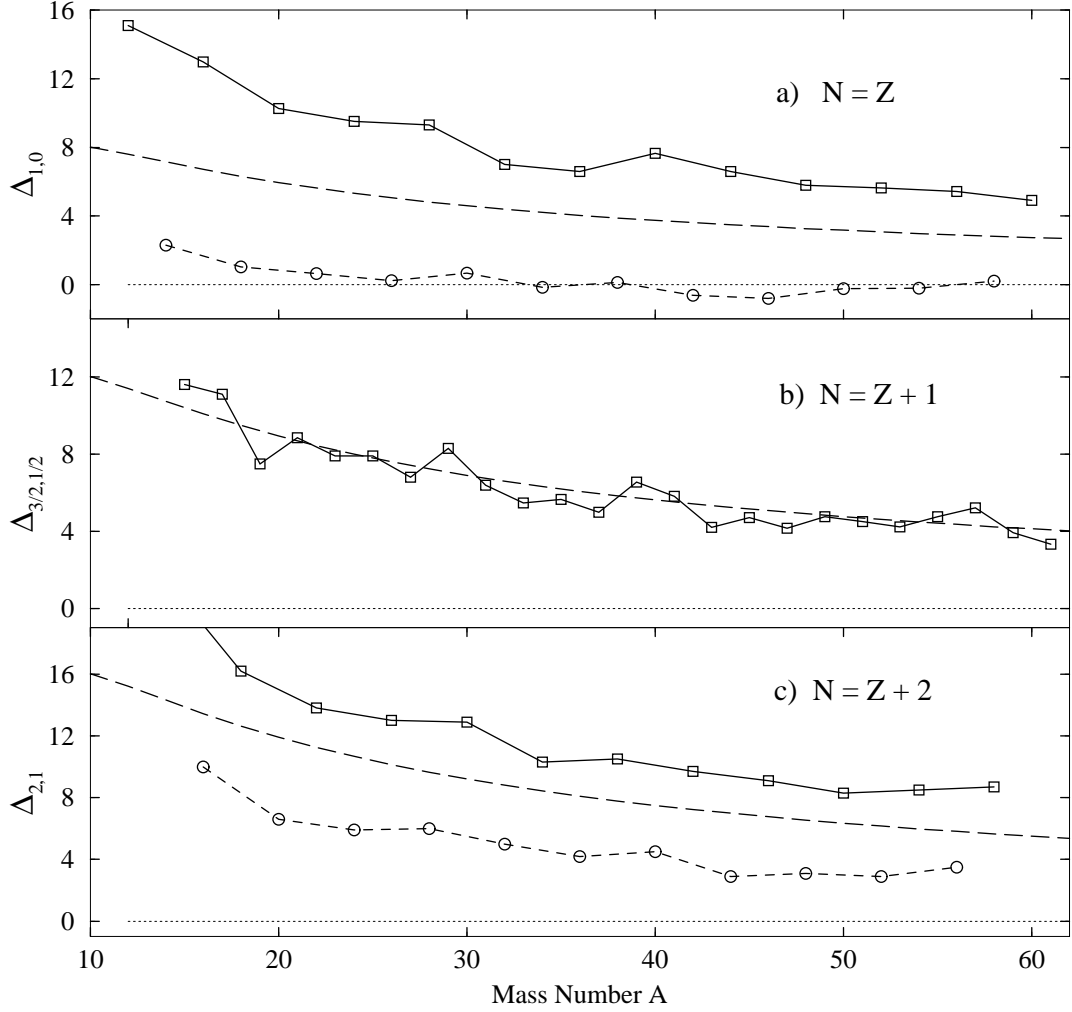}}
\vspace{1.5cm}
\caption{Experimental excitation energies $\Delta_{T',T}$
(in MeV) of the lowest $T' = |N-Z|/2 +1$ state with respect
to the lowest $T = |N-Z|/2$ state. 
Panel a) is for the $N = Z$ nuclei, b) is for the $N = Z +1$
nuclei, and c) for the $N = Z + 2$ nuclei. 
In panels a) and c)  the squares connected
by full lines to lead the eye
are for the even-even nuclei, and the circles connected by short dashed
lines are for the odd-odd nuclei. 
In all three panels the long dashed line is the symmetry
energy difference defined in Eq. (\protect{\ref{e:sym}}). 
The thin dotted lines indicate zero excitation energy. }
\label{f:exc}
\end{center}
\end{figure}

Consider the ground state of a nucleus with $A = N + Z$ nucleons, 
with $A$ even.
This state can be a condensate of 
only $J^{\pi} = 0^+$ isovector pairs if $T = | N - Z |/2$ for 
even-even nuclei, and if $T = | N - Z |/2 + 1$ 
for the odd-odd nuclei. In other words,
states with  $T = | N - Z |/2 + 1$ in even-even nuclei must 
contain at least one broken
pair, and similarly states with $T = | N - Z |/2$ in 
odd-odd nuclei must also contain at least
one broken pair. In contrast, in the odd-$A$ nuclei  
one can have states with both $T = |N - Z|/2$ and
$T = | N - Z |/2 + 1$  all having just one unpaired nucleon. Based on these
considerations one can understand the trends shown in Fig. \ref{f:exc}.
The quantities $\Delta_{T',T}$ plotted in Fig. \ref{f:exc} are the
excitation energies of the lowest states of isospin $T'$ with respect
to the states with isospin $T$. Analogous figures can be found
in Refs. \cite{Janecke,Baz}.

Let us discuss first the middle panel b) in Fig. \ref{f:exc}
for the odd-$A$ $N = Z + 1$ nuclei. Disregarding
the relatively small shell effects, 
the excitation energies of the lowest $T = 3/2$
state form a smooth monotonic curve. Indeed since, as I argued above, pairing
should be the same for both isospins, the excitation energy should be simply
related to the symmetry energy. In fact, the dashed line in panel b), based
on the symmetry energy formula fitted in Ref. \cite{Zuker},
\begin{equation}
E_{sym} = \frac{T(T+1)}{A} 
\left[ ~ 134.4 - 203.6 \frac{1}{A^{1/3}} ~ \right] ~{\rm (MeV)} ~,
\label{e:sym}
\end{equation}
approximates the data very well.
Note that this formula results from the fit to masses of 1751 nuclei 
(together with the other parameters of the set $6p$) in Ref. \cite{Zuker}. 
Here I confirm its validity using just few excitation energies, 
and in fact it should be possible to fit it independently from the data in 
the panel b)
alone. In particular, the sign and magnitude of the correction (surface)
term is verified.

However, the even-$A$ nuclei in panels a) and c) are quite different. There,
the excitation energies form $two$ curves
as already noted in Refs.\cite{Janecke,Baz}.
In the upper one for the even-even nuclei
the symmetry energy and pairing add, and in the lower one for the 
odd-odd nuclei the excitation energy is the difference between the
symmetry energy effect and pairing, i.e. they act 
against each other. That this is the correct interpretation
of the data is verified  by the fact that the long dashed 
lines in panels a) and c),
based on the symmetry energy, Eq. (\ref{e:sym}), bisect the two curves. (These
lines are not exactly in the middle of the two lines, 
in particular in c). I will comment
on the possible reason for this later.)  Thus, while the average of the
even-even and odd-odd curves in Fig. \ref{f:exc} represents the symmetry
energy, their difference represents twice the energy needed to break an
isovector pair.
Fig. \ref{f:exc} is based on excitation
energies as listed in Table of Isotopes \cite{Table}. In few cases, however,
the isospin assignment is not available. In those cases I used the mass of the
corresponding analog nucleus, corrected for the neutron-proton mass
difference and the Coulomb energy also based on Ref. \cite{Zuker}
\begin{equation}
E_{Coul} = -0.699 \frac{Z(Z-1)}{A^{1/3}} 
\left[ 1 - \frac{0.76}{[Z(Z-1)]^{1/3}} \right] ~{\rm (MeV)}~.
\label{e:coul}
\end{equation}
In cases where it can be checked this formula gives quite good agreement 
with the excitation energies displayed in Fig. \ref{f:exc}.
However, the Coulomb energy is otherwise irrelevant for the further
discussion.

Several features of Fig. \ref{f:exc} deserve special comment. 
First, one can see
that in the $N = Z$ odd-odd nuclei  the 
lowest $T = 1$ and $T = 0$ states are almost
degenerate. There is very little difference 
in this respect between the $sd$ shell nuclei ($A < 40$),
where $T = 0$ is the ground state,
and the heavier $pf$ shell nuclei where the $T = 1$ is 
usually the ground state. 
The small difference seen in Fig.  \ref{f:exc}
can be obviously correlated with the curvature of the symmetry
energy line; it has apparently little to do with the strength
of the isovector and isoscalar pairing. 
This degeneracy of the $T$ = 0 and 1 states
means that, remarkably, in these nuclei the
symmetry and pairing energies are equal in magnitude,
i.e. {\it almost exactly cancel} each other.  
Based on the figure one would expect that the $T = 1$ and $T = 0$ 
states will remain close to each other
also in the heavier $N = Z$ odd-odd nuclei.

One can understand also that the odd-odd $N = Z$ nuclei are the only ones
known that violate the rule that the ground state isospin is $T = |N - Z|/2$. 
In  essentially all other 
nuclei the symmetry energy is stronger than pairing, such as in c) and 
even more
so in nuclei with larger $|N - Z|$ and/or ground
state isospin. In Ref.\cite{Janecke} the symmetry and pairing energies
were fitted from the quantities $\Delta_{T',T}$, with the shell effects
included. Based on these fits, J\"{a}necke \cite{Janecke} predicted that the
isospin ``inversion'' should also occur in other 
odd-odd very proton rich nuclei,
e.g. for $T_z = \pm 1$ for 108$\le A \le$ 124 as well as for $A \ge$ 192,
and for $T_z = \pm 2$ for $A \ge$ 290. Most of such nuclei are,
however, beyond the proton drip line.

Further, it is clear that the symmetry energy should depend on the isospin $T$
and not just on the 
square of the neutron excess $(N - Z)^2$ as in the usual liquid drop
formula. (In most heavier nuclei, however, where the liquid drop
formula is used, $T=|N-Z|/2$ is much larger than unity and the difference
between $T(T+1)$ and $(N-Z)^2/4$ plays very little role.)
The parametrization in Eq. (\ref{e:sym}), containing $T(T+1)$, is
motivated by the charge independence of the nuclear force. But as
a phenomenological parametrization the isospin dependence 
$T(T + a)$ with $a \ne 1$ is also possible. 
In fact, the empirical fits in Ref.\cite{Janecke} show a clear preference
for $a=1$; the fits with either $a=0$ as in the usual liquid drop
formula, or with $a=4$ as in the Wigner SU(4) symmetry are 
disfavored. Similar conclusion, namely that $a \simeq 1$, 
has been reached in Ref.\cite{Axel}.

The presence of a term linear in $T = |N - Z|/2$ has a 
special relevance for the so-called
Wigner energy\footnote{Prof. A. Zuker stressed this point in a private
communication to the author}. The term, first introduced by
Wigner \cite{Wignero}, is used for the additional binding energy in the
semi-empirical mass formulae,
\begin{equation}
E_W = W(A) | N - Z | + d(A) \pi_{np}\delta_{NZ} ~,
\label{e:Wig}
\end{equation}
where $W(A)$ is a smooth function of $A$ describing the
magnitude of the effect. (The quantity 
$\pi_{np} = 1$ for odd-odd nuclei and vanishes otherwise. 
The $d(A)$ term, relevant only in the odd-odd $N = Z$ nuclei is 
not discussed further here.)
The empirical fit to the first term in Eq. (\ref{e:Wig})
gives $E_W \simeq 47|N-Z|/A$ MeV \cite{Wigner}.
The symmetry energy formula, Eq.(\ref{e:sym}) contains
a term  (for $A \simeq 40$) $37|N-Z|/A$ MeV not very far from
the empirical Wigner energy value. This finding is in agreement
with  Ref.\cite{Poves} where it was shown that the experimental
magnitude of the 
Wigner term is reduced substantially if one uses $T(T+1)$,
instead of the more common $(N - Z)^2$ for the symmetry energy. 

While reducing the need for the extra Wigner energy term, Eq. (\ref{e:sym})
does not eliminate it completely. This could be achieved, perhaps, by
using the form $T(T+a), ~a > 1$ for the symmetry energy. Such a choice
would also move the middle curve in panel c) of Fig. \ref{f:exc} towards
the average of the curves for even-even and odd-odd nuclei. 
Thus, there is an indication that the ``best'' semi-empirical symmetry
energy formula would have $a > 1$. No attempt has been made
to do such a fit here; a more physical approach would relate the
Wigner energy to the various components of the neutron-proton
force such as in Ref. \cite{Wigner}.

What is then the magnitude of the pairing gap in the $N \simeq Z$
nuclei? The usual definition \cite{BM} relates the gap to the binding energy
difference of the given even-$A$ nucleus to its odd-$A$ neighbors
eliminating other smooth trends.
For example, for the neutron gap one is supposed to use
\begin{equation}
\Delta_{nn} (N,Z) = \frac{1}{4} \left[ B(N-2,Z) - 3B(N-1,Z)
+ 3B(N,Z) - B(N+1,Z) \right] ~.
\end{equation}
However, in $N \simeq Z$ nuclei such definition is not really applicable, since
not only the number of the $nn$ pairs changes, but also the number of
$np$ pairs.

\begin{figure}
\begin{center}
\leavevmode 
\rotate[r]{\epsfxsize=0.7\textwidth 
\epsffile{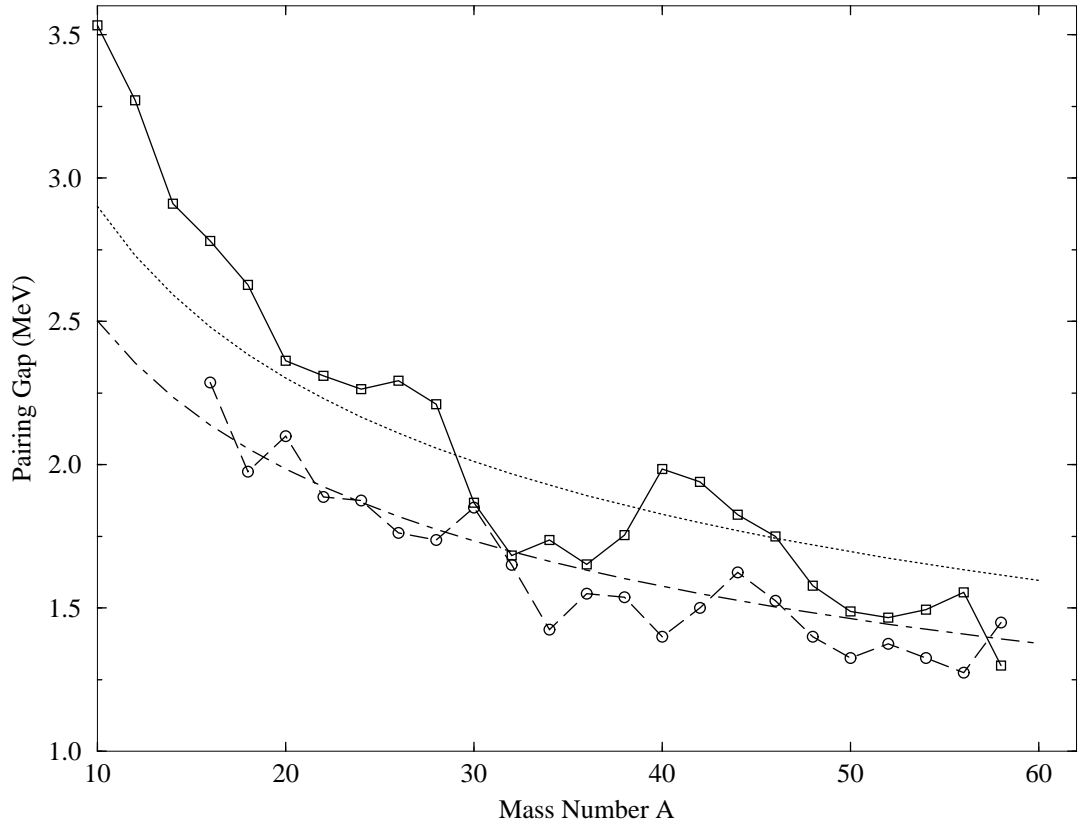}}
\vspace{1.5cm}
\caption{Pairing gaps for the $N = Z$ nuclei (squares connected
by the solid line) and for the $N = Z + 2$ nuclei (circles connected
by the long dashed line). The smooth fits are $6.24/A^{1/3}$ (dotted
line for $N = Z$) and $5.39/A^{1/3}$ 
(dot-and-dashed line for $N = Z + 2$). All gaps are in MeV
and were extracted by the method described in the text.
}
\label{f:pair}
\end{center}
\end{figure}

We can use, however, the excitation energy of the 
lowest $T_> = |N - Z|/2 +1$ state with respect 
to the lowest $T_< = |N - Z|/2$ one, as shown in Fig. \ref{f:exc}, for this
purpose. Since the lines are there just to guide the eye, one must
use some form of interpolation, however. In Fig. \ref{f:pair}, I use the
simplest averaging in order to obtain the gap for an even-even nucleus,
\begin{eqnarray}
\Delta_{ee} (N,Z) & = & \frac{1}{4} 
\{ E_{ee} (N,Z,T_>) - E_{ee}(N,Z,T_<)  \nonumber \\
& & - \frac{1}{2} [ E_{oo} (N-1,Z-1,T_>) + E_{oo} (N+1,Z+1,T_>)  \nonumber \\
& & - E_{oo} (N-1,Z-1,T_<) - E_{oo} (N+1,Z+1,T_<) ]  \} ~.
\label{e:gap}
\end{eqnarray}
An obvious modification is used for the odd-odd nuclei. 

The results shown in Fig. \ref{f:pair} show, first of all, 
that the pairing gaps for a given
$N - Z$ form a more or less continuous curves, 
with however clearly visible shell
effects. The gaps defined using Eq. (\ref{e:gap}) 
do not vanish at magic numbers,
showing that the odd-even staggering is indeed 
a more general feature of nuclear
spectra, as suggested in Ref. \cite{Jacek}. The smooth trends can be fitted as
$6.24/A^{1/3}$ MeV for the $N = Z$ nuclei and $5.39/A^{1/3}$ MeV for the
$N = Z + 2$ nuclei. (The curve for the $N = Z + 4$ nuclei, not shown, is very
close to the one for $N = Z + 2$.) In Ref. \cite{Zuker} the general pairing gap
was fitted as $5.18/A^{1/3}$ MeV, quite close to our fit for $N > Z$. 
The larger gap
for $N = Z$ reflects the gain in pairing due to stronger $np$ correlations.

In conclusion, following Ref. \cite{Janecke}
we have seen that in the $N \simeq Z$ 
nuclei the effects of pairing
and symmetry energy are closely related. In particular, in the odd-odd 
$N = Z$ these two effect are essentially equal in magnitude, 
and their cancellation causes the
near degeneracy of the lowest $T = 1$ and $T = 0$ states. We have
also shown that the symmetry energy extracted from the excitation
energies of the $T_>$ states agrees with the global fit. 
Moreover, the considerations presented here show that the
proper parametrization of the symmetry energy must be 
a quadratic function of the isospin $T$ of the form $T(T+a)$
(and not $(N - Z)^2$).
Based on isospin symmetry, one would naturally choose
$a = 1$, but the purely empirical value, which would at the same
time explain most of the Wigner energy, favors somewhat larger 
value of $a$. Finally, 
the pairing gaps extracted from these data by the procedure in 
Eq. (\ref{e:gap}) agree with the usual definition and global fit
for the $N > Z$ nuclei. We find evidence for about 20\% increased gaps
in the $N = Z$ nuclei.

\vspace{1cm}

Discussion with Prof. Andres Zuker on the symmetry and Wigner
energies is gratefully acknowledged.
This work was supported  by the U. S. Department of Energy under
grant No. DE-FG03-88ER-40397.

\end{document}